\documentclass[pra,twocolumn,showpacs,preprintnumbers,amsmath,amssymb]{revtex4}
\usepackage{graphicx}
\usepackage{dcolumn}
\usepackage{bm}
\usepackage{latexsym,epsfig}
\usepackage{textcomp}
\usepackage{color}

\begin{document}

\title{Hardcore bosons in a zig-zag optical superlattice}
\author{Arya Dhar$^1$, Tapan Mishra$^2$, Ramesh V. Pai$^3$, Subroto Mukerjee$^{4,5}$, B. P. Das$^1$}
\affiliation{$^1$ Indian Institute of Astrophysics, Bangalore, 560034, India}
\affiliation{$^2$International Center for Theoretical Sciences, Tata Institute of Fundamental Research,
Bangalore, 560 012, India}
\affiliation{$^3$ Department of Physics, Goa University, Taleigao Plateau, Goa 403 206, India}
\affiliation{$^4$ Department of Physics, Indian Institute of Science,
Bangalore, 560 012, India}
\affiliation{$^5$ Centre for Quantum Information and Quantum Computing (CQIQC), Indian Institute of Science,
Bangalore, 560 012, India}

\date{\today}

\begin{abstract}
 We study a system of hard-core bosons at half-filling in a one-dimensional
 optical superlattice. The bosons are allowed to hop to nearest and next-nearest neighbor sites. We
 obtain the ground state phase diagram as a function of microscopic parameters using the finite-size
 density matrix renormalization group (FS-DMRG) method. Depending on the
sign of the next-nearest neighbor hopping and the strength of the
superlattice potential the system exhibits three different phases,
namely the bond-order (BO) solid, the superlattice induced Mott
insulator (SLMI) and the superfluid (SF) phase. When the signs of both
hopping amplitudes are the same (the ``unfrustrated'' case), the
system undergoes a transition from the SF to the SLMI at a non-zero
value of the superlattice potential. On the other hand, when the two
amplitudes differ in sign  (the ``frustrated'' case), the SF is
unstable to switching on a superlattice potential and also exists
only up to a finite value of the next nearest neighbor hopping. This
part of the phase diagram is dominated by the BO phase which breaks
translation symmetry spontaneously even in the absence of the
superlattice potential and can thus be characterized by a bond order
parameter. The transition from BO to SLMI appears to be first order.
\end{abstract}


\maketitle

\section{Introduction}
 Ultracold atoms provide a unique opportunity to investigate a wide range of phenomena, especially
 in low dimensions where quantum fluctuations play a dominant role~\cite{rigol_rmp}. Because of the exquisite control 
 and precision possible experimentally, they offer nearly perfect realizations of various model condensed
 matter systems. The seminal paper by Jaksch et al
 ~\cite{jaksch} which had predicted the quantum phase transition from Mott insulator to superfluid
 phase in the Bose-Hubbard model, paved the way for the first experimental observation of this
 transition in an optical lattice by Greiner et al~\cite{greiner}.
 Current experimental techniques have successfully created various lattice geometries using
 proper arrangements of laser beams, such as optical superlattices~\cite{bloch,blochsuperlat}, triangular~\cite{becker}
  and Kagome lattices~\cite{jo}. Such a diverse class of lattice systems
 gives us the opportunity to study a variety of models which might also be geometrically frustrated. The
 interatomic interactions can be controlled to a high degree of accuracy with the help of
 Feshbach resonances~\cite{chin}. Recent developments in shaking techniques have enabled
 experimentalists to modify the value and sign of the intersite hopping~\cite{holthaus,arimondo}
 thus opening up possibilities to investigate frustrated systems of bosonic lattice gases~\cite{struck}. {\text


 \begin{figure}[!t]
  \centering
  \psfig{file=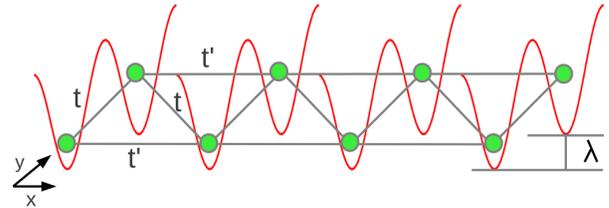, scale=0.30,clip=true}
  \caption{(Colour online) Schematic diagram for a zig-zag optical superlattice
  with nearest (t) and next-nearest neighbor ($t'$) hopping. $\lambda$ is the optical superlattice potential.}
  \label{fig:model}
 \end{figure}

 Earlier works on ultracold bosons in optical superlattice have shown
 the existence of phases with density wave-like configurations~\cite{roth,rousseau,rigol,danshita,ying,shuchen}.
 Later studies of soft-core bosons
 in optical superlattices~\cite{aryadmrg, aryamft} termed these phases as superlattice induced Mott insulator(SLMI) phases. Recent studies
 on models dealing with the interplay
 between frustration imposed by geometry and interactions have revealed rich physics with a variety of
 novel phases being exhibited~\cite{aryabose1,aryabose2,santos,tapan}.

In this paper, we analyze a system of hardcore bosons in a $1D$ superlattice with nearest and 
next-nearest neighbor hoppings. The superlattice potential creates an energy offset in alternate sites. 
This model is equivalent to a zig-zag superlattice as shown in Fig.~\ref{fig:model}. The nearest and the 
next-nearest neighbor hoppings are equivalent to the hoppings between the legs and within the legs of the zig-zag 
lattice respectively. The energy offset can be introduced by applying a constant electric field in the y-direction 
as shown in Fig.~\ref{fig:model}.  
In such a situation the system can
be described by the Hamiltonian given by
\begin{eqnarray}
  H&=&-t\sum_i{(a_i^{\dagger}a_{i+1}+h.c.)}-t'\sum_i{(a_i^{\dagger}a_{i+2}+h.c.)}\nonumber \\
  &&+\sum_i{\lambda_in_i}
  \label{eq:model}
 \end{eqnarray}
where $a_i^{\dagger}$ and $a_i^{\phantom \dagger}$ are creation and annihilation operators
for hard core bosons
at site $i$, and $n_i=a_i^{\dagger}a_i^{\phantom \dagger}$ is the boson number operator
at site $i$. Here $t$ and $t'$ are the hopping amplitudes for tunneling to a neighboring site
 and a next-nearest neighbor site respectively and
 $\lambda_i$ is the superlattice potential.
 In the present work, we have considered a two-period
 superlattice, with $\lambda_i=\lambda$ for odd $i$ and zero for even $i$. We assume that the values of $t'$ 
 from even to even sites and from odd to odd sites are equal in magnitude. In other words the hopping amplitudes 
 along the legs of the zig-zag lattice are the same. A similar assumption has been made in an earlier work on square ladder~\cite{danshita} 
 and also in a recent experiment in a sqaure lattice ~\cite{blochsuperlat}.
 We study the system for a wide range of $t'$ and $\lambda$ and we fix the energy scale in the units
 of $t$ by taking the value of $t=1$. 

 \begin{figure}
  \centering
  \psfig{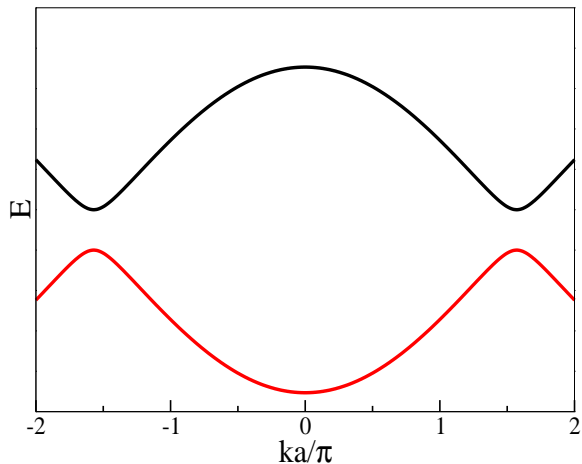}
  \caption{(Color online) Dispersion relation as computed from the Eq.~(\ref{Eq:dispersion}).}
  \label{fig:dispersion}
 \end{figure}

When $t'=0$ the above model can be mapped onto a non-interacting
model of spinless fermions. Using the Jordan-Wigner
transformation~\cite{JW}
\begin{eqnarray}
a^{\dag}_i=f^{\dagger}_i \prod^{i-1}_{\beta=1}e^{-i\pi f^{\dagger}f} , \hspace{0.5cm}
a_i=\prod^{i-1}_{\beta=1}e^{-i\pi f^{\dagger}f} f^{\dagger}_i
\end{eqnarray}
 Eq.~(\ref{eq:model}) can be mapped to
\begin{eqnarray}
  H&=&-t\sum_i{(f_i^{\dagger}f_{i+1}+h.c.)}+\sum_i{\lambda_if^{\dagger}_i f_i}
  \label{eq:spinless}
 \end{eqnarray}
where $f^{\dagger}_i$ and $f_i$ are the creation and annihilation
operators for the spinless fermions and $f^{\dagger}_i f_i$ is the
fermion number operator. The single particle eigenstates of the
Hamiltonian given by Eq.~(\ref{eq:spinless}) can be obtained
exactly. There are two bands arising from the fact that the
translational symmetry of the lattice has been broken by the
superlattice potential. The energy spectra of the two bands are given
by
\begin{equation}
E_{\pm} (k)= \frac{\lambda \pm \sqrt{\lambda^2+\left[4t\cos(ka)\right]^2}}{2},
\label{Eq:dispersion}
\end{equation}
where $a$ is the lattice spacing and $k$, the crystal momentum that
runs from $-\frac{\pi}{2a}$ to $\frac{\pi}{2a}$. A plot of these
spectra is shown in Fig.~\ref{fig:dispersion}. From this figure and
Eq.~(\ref{Eq:dispersion}), it can be seen that there is a gap equal
to $\lambda$ at half filling. Turning on $t'$ augments the effect of
$t$ if they have the same sign yielding a superfluid as we show
below. For the opposite sign of $t'$ the system is frustrated and we
find that this frustration coupled with the superlattice potential
prevents superfluidity from occurring. For either sign of $t'$, the
model can no longer be mapped onto one of non-interacting spinless
fermions and thus we have to take recourse to numerics to study it.
We do this via a state-of-the art density matrix renormalization
group (DMRG) method~\cite{white}.

The remaining part of the paper is organized as follows. In Section II, we
 outline the method of calculation we have used, followed by a presentation of results in Section III and a summary of our conclusions in section IV.

 \section{Method of calculation}



We study the ground state properties of the model described by
Eq.~(\ref{eq:model}) using the finite-size DMRG method with open
boundary conditions~\cite{white_92,schollwock_review_05} which is
best suited to (quasi-)one-dimensional
problems~\cite{schollwock_review_05}. For our calculations we study
system sizes up to 300 sites and retain $128$ density matrix
eigenstates with the weight of the discarded states in the density
matrix being less than $10^{-6}$.

In order to obtain the ground state phase diagram we calculate several physical quantities of interest.
Some of these quantities have been calculated by us using the DMRG method to study
related models~\cite{tapan,tapan_tvvp}.
To separate out the gapped and gapless phases we calculate the single particle excitation gap given by
\begin{equation}
G_L=E(L,N+1)+E(L,N-1)-2E(L,N).
\label{eq:gap}
\end{equation}
In Eq.~(\ref{eq:gap}),
$E(L,N)$ is the ground-state energy of a system with $L$ sites and $N$ bosons.
To identify the BO phase we compute the bond order parameter given by
  \begin{eqnarray}
   O_{BO}=\frac{1}{L}\sum_i (-1)^i B_i
   \label{eq:boopeq}
  \end{eqnarray}
where $B_i=\langle b^{\dagger}_i b_{i+1}+b^{\dagger}_{i+1}b_i \rangle$ is the bond energy.
The presence of the SLMI phase can be determined through a calculation of the structure factor obtained by taking the Fourier transform of the density-density correlation function
  \begin{eqnarray}
   S(k)=\frac{1}{L^2}\sum_{i,j}{e^{i(i-j)k}\langle n_i^{\dagger} n_j \rangle} 
   \label{eq:strucfacteq}
  \end{eqnarray}
We also calculate the momentum distribution function given by
  \begin{eqnarray}
   n(k)=\frac{1}{L}\sum_{i,j}{e^{i(i-j)k}\langle a_i^{\dagger} a_j \rangle }
   \label{eq:momentumeq}
  \end{eqnarray}

Before discussing the results obtained from our FS-DMRG calculation
we mention that  we have verified that our numerical method gives us
very accurate results in the analytically tractable case $t'=0$. We
compute the single particle excitation gap using Eq.~(\ref{eq:gap})
for this case, which from the analysis of the previous section is
equal to $\lambda$. Our numerics give us a gap which is within
$0.01$ percent for various values of $\lambda$. We thus believe that
our FS-DMRG calculation yields very accurate results even for $t'
\neq 0$.



 \section{Results}
 \begin{figure}[!t]
  \centering
  \psfig{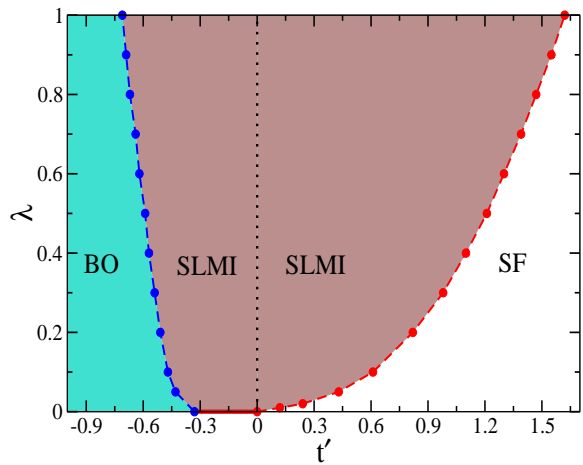}
  \caption{(Color online) Phase diagram for a system of hard-core bosons with nearest-neighbor ($t=1$) and next nearest-neighbor ($t'$)
  hopping
  in an optical superlattice with potential $\lambda$, at a filling factor of half.}
  \label{fig:phase-diagram}
 \end{figure}
 We now discuss the results of the present work.  Before
 we give the results of our numerical calculations in detail,  we
 present arguments for the general structure of the phase diagram.
 As discussed in section I, when $t'=0$, the system has a gap for any finite value of $\lambda$.
 The presence of the gap would make the state robust to perturbations due to $t'$,
 till they get to be roughly of the order of the gap. The SLMI phase is the adiabatic continuation of the $t'=0$ gapped phase for $t' \neq 0$.
The gap at $t'=0$ increases with increasing $\lambda$. As a result,
the extent of the SLMI phase along the $t'$ axis will be larger as
$\lambda$ increases. This can indeed be seen in the phase diagram
obtained from our numerical calculations shown in
Fig.~\ref{fig:phase-diagram}, where we have set the energy scale by
$t=1$.

We now consider the phases that arise at large values of $|t'|$ after the SLMI phase has disappeared.
 We note that when $\lambda=0$, our model as described in Eq.~(\ref{eq:model}) is same as hardcore bosons hopping on a triangular ladder. For $t'>0$,
 we have determined from DMRG calculations that the ground state is always a superfluid and we find that this state
 is not immediately destroyed by a non-zero value of $\lambda$. There is thus a phase boundary between the SLMI
 and SF as shown in Fig.~\ref{fig:phase-diagram} in the $\lambda- t'$ plane. For $t'< 0$ and $\lambda=0$,
 the system is ``frustrated'' and it is known that the superfluid does not persist up to arbitrarily large
 values of $|t'|$~\cite{tapan}. The superfluid persists up to $|t'| \sim 0.33$ after which a gapped bond
 ordered (BO) phase forms through a Berezinski-Kosterlitz-Thouless (BKT) type transition. A physical
 picture of the BO phase can be obtained by studying the model at the exactly solvable point $|t'|=0.5$,
 where it can be mapped onto an $XY$ version of the spin 1/2 Majumdar-Ghosh spin chain~\cite{majumdar}.
 The BO phase is a valence bond solid, where the valence bonds are of the type
 $\frac{1}{\sqrt{2}} \left(|0 \rangle_i |1 \rangle_{i+1}+ |0 \rangle_{i+1} |1 \rangle_i \right)$
 for adjacent sites $i$ and $i+1$. Note, that this state spontaneously breaks translational symmetry
 even for $\lambda=0$. Since the BO phase is gapped, we expect it to be stable up to a critical value of $\lambda$
 after which the SLMI phase emerges. This is indeed the case as can be seen from Fig.~\ref{fig:phase-diagram}.
 The SF phase that exists below $|t'| \sim 0.33$, however, appears to be unstable to the introduction of $\lambda$
 in contrast to the SF phase for $t'>0$. It should be noted that though both the SLMI and the BO phases are gapped
 and not translationally invariant, they are fundamentally different in that the BO phase has a non-zero BO order
 parameter given by Eq.~(\ref{eq:boopeq}), which is zero for the SLMI phase. All three phases have a non-zero structure
 factor $S(k=\pi)$ given by Eq.~(\ref{eq:strucfacteq}) when $\lambda \neq 0$ since translational symmetry is being broken by hand.
 However, as we will show there is a kink in the structure factor as a function of $t'$ at the boundary between the BO and SLMI.


\subsection{Positive $t'$ case}
As argued above, we expect a transition between the SLMI and SF phases for $t' > 0$.
Since this transition is from a gapped to gapless phase, it is determined numerically by calculating the single particle
 \begin{figure}[t]
  \centering
  \psfig{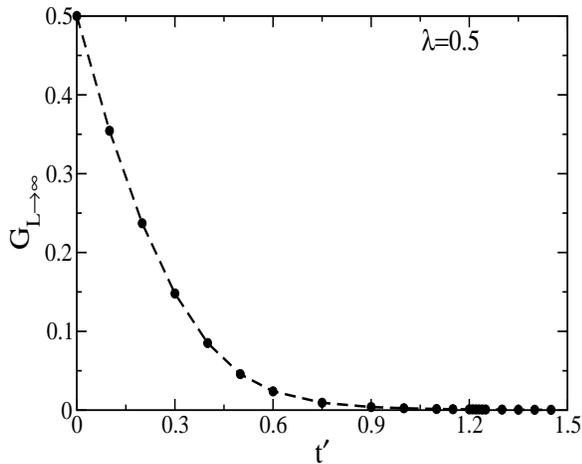}
  \caption{Thermodynamic values of $G$ plotted against $t'$ for $\lambda=0.5$ to
  locate the transition point.}
  \label{fig:thermogap}
 \end{figure}
excitation gap as defined in Eq.~(\ref{eq:gap}). We perform a finite
size scaling of the gap $G_L$ by fitting a quadratic polynomial in
$1/L$ and extrapolating it to $L\rightarrow \infty$ to get the
thermodynamic limit values of $G$. For the fitting we consider
fairly large system sizes i.e. from $L=100$ to $L=300$. The
extrapolated values of $G_L$ as a function of $t'$ for $\lambda=0.5$
are shown in Fig.~\ref{fig:thermogap} which clearly shows a
transition from a gapped to gapless phase. The gap
appears to close slowly as the value of $t'$ approaches the critical value at $1.21$. 
In Fig.~\ref{fig:gapextrapolate} we show the finite size scaling of the gap $G_L$ for different values of $t'$.
It can be seen that the fitting functions gradually go to zero as the
transition point approaches. We estimate the critical point by noting that the extrapolated gap $G_{L \rightarrow \infty}$
appears to stabilize to a value that is less than $10^{-3}$. Since, we expect the gap to approach to zero in the true
thermodynamic limit in the SF phase, we obtain the phase boundary by using the criterion that for
$G_{L\rightarrow \infty}$ less than $10^{-3}$ the system is gapless.
 \begin{figure}[b]
 \centering
 \psfig{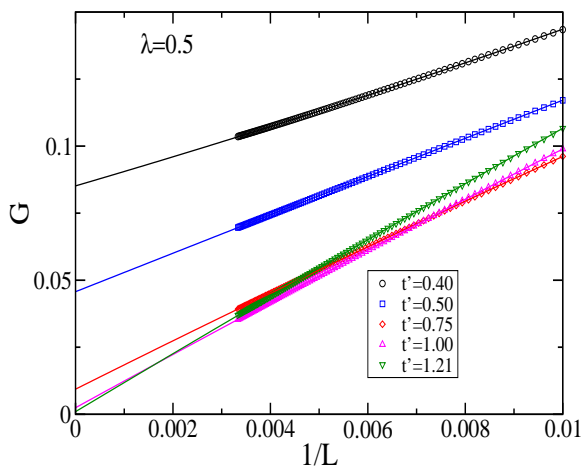}
 \caption{(Color online) Gap, $G$, plotted against $1/L$, along with the extrapolation for different values of $t'$
 for $\lambda=0.5$}
 \label{fig:gapextrapolate}
 \end{figure}

 \subsection{Negative $t'$ case}

 To determine the BO phase, we calculate the order parameter $B_i$ and plot it as a function of $i$ for different values of $t'$.
 This is done for $\lambda=0.5$ in Fig.~\ref{fig:avgebij}.
 \begin{figure}[!t]
  \centering
  \psfig{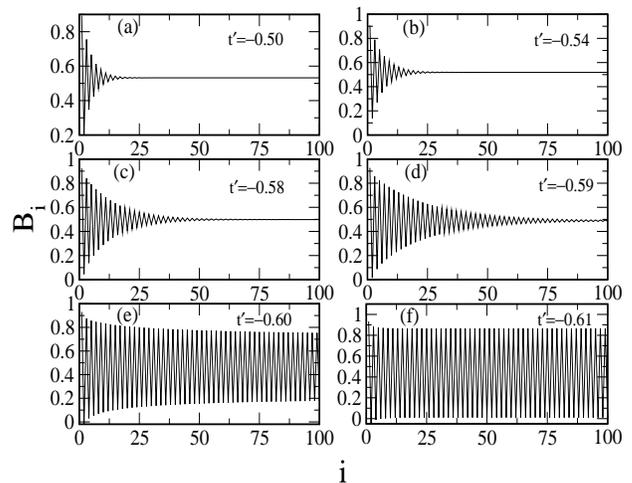}
  \caption{$B_{i}$ is plotted against $i$ for $\lambda=0.5$ for different values of $t'$.}
  \label{fig:avgebij}
 \end{figure}
It can be clearly seen that for small values of $|t'|$,
 there is an exponential decay in $B_{i}$ (Figs.~\ref{fig:avgebij}(a)-(d)). However, in Fig.~\ref{fig:avgebij}(e),
 we observe the emergence of long-range bond oscillations. In Figs.~\ref{fig:avgebij}(e) and (f), there are distinct
 oscillations throughout the lattice. This indicates the presence of the BO phase at higher negative
 values of $t'$. In order to locate the transition to the BO phase we compute the bond-order parameter
defined in Eq.~(\ref{eq:boopeq}).

 \begin{figure}[b]
  \centering
  \psfig{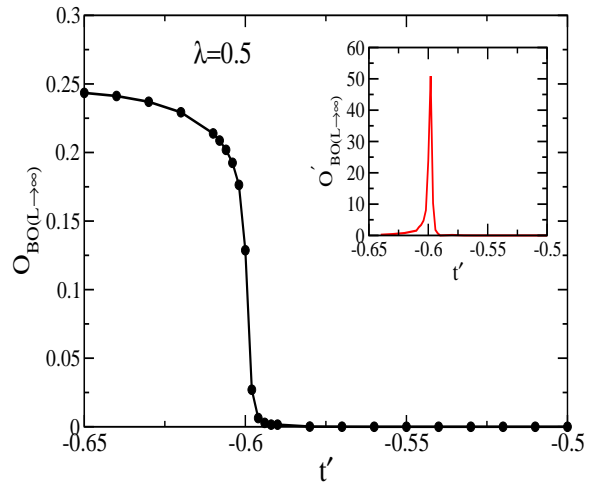}
  \caption{(Colour online) Plot of thermodynamic values of the $O_{BO}$ against $t'$
  for $\lambda=0.5$. A discrete jump in the values can be observed around the
  transition point. Inset: The first derivative $O'_{BO(L\rightarrow \infty)}$ showing a peak at the transition point from
  SLMI to BO phase.}
  \label{fig:thermoboop}
 \end{figure}
 We plot the
 thermodynamic values of the $O_{BO}$ obtained from a third order polynomial extrapolation, as shown
 in Fig.~\ref{fig:thermoboop}. The BO phase is expected to have a finite $O_{BO}$ whereas it will be zero
 in the SLMI phase. As can be clearly seen from the figure, a discrete jump
in its value is noticed, which is further supported by a sharp peak in the first derivative as shown in the inset.
 This clearly signifies a phase transition
 to the BO phase from the SLMI phase. The transition is located by taking the derivative maximum of the
 $O_{BO(L\rightarrow \infty)}$ given by $O'_{BO(L\rightarrow \infty)}=dO_{BO(L\rightarrow \infty)}/dt'$,
 as shown in the inset of Fig.~\ref{fig:thermoboop}.

 The bond order parameter calculation to locate the SLMI-BO transition critical point is complemented by the scaling of the single particle excitation gap $G_L$. In
  Fig.~\ref{fig:gapvstp} we plot $G_L$ as a function of $t'$ for different lengths and for $L\rightarrow \infty$ obtained by extrapolation.
It can be seen that the system is always gapped along the $\lambda$ axis. The gap decreases as the
critical point approaches and remains finite and then increases again. The minimum shifts towards the actual critical point
for larger lengths. Extrapolating to the thermodynamic limit we find the minimum to occur at the critical point $t'=-0.604$
as obtained from the $O_{BO}$ scaling. Note, however, that the gap
does not go to zero at the transition indicating a first order
transition consistent with the jump of the BO order parameter.
 \begin{figure}[t]
  \centering
  \psfig{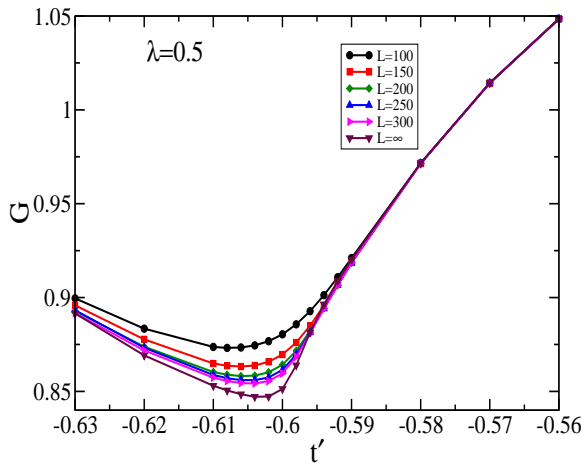}
  \caption{(Color online) Energy gap, G plotted for different lengths against $t'$. The minimum
  of the $L\rightarrow\infty$ curve implies the critical point.}
  \label{fig:gapvstp}
 \end{figure}
 \begin{figure}[b]
  \centering
  \psfig{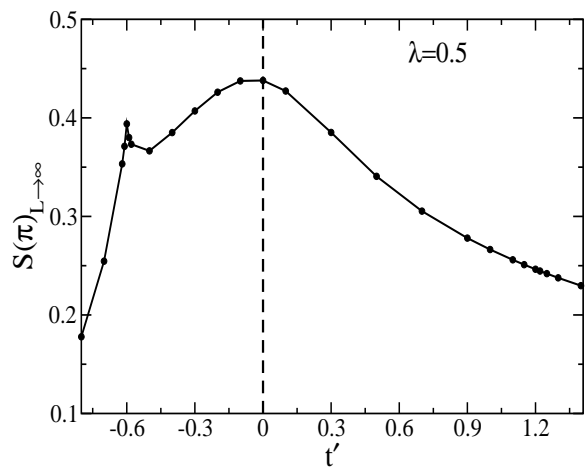}
  \caption{Thermodynamic values of $S(\pi)$ is plotted for the entire range of $t'$ for
  $\lambda=0.5$.}
  \label{fig:thermospi}
 \end{figure}

The imposition of the superlattice potential $\lambda$ will cause a density modulation of the
type [...1 0 1 0 ...] in all the three phases. The structure factor, $S(k)$, as defined in
 Eq.~(\ref{eq:strucfacteq}) will show finite peaks at $k= \pm \pi$, whose heights are large in the
 SLMI phase and smaller in the BO and the SF phases. The thermodynamic value of $S(\pi)$ is plotted for $\lambda=0.5$
 in Fig.~\ref{fig:thermospi}. In the BO phase, $S(\pi)$ is small and increases steadily with
 decreasing $|t'|$. At the transition point between BO and SLMI, it has a kink and then increases
 gradually as the value of $t'$ approaches $0$. A similar plot for a smaller value of $\lambda=0.05$
 is shown in Fig.~\ref{fig:thermospi-lambda0p05} showing a similar peak in the negative $t'$ region at the
 transition point. In the positive $t'$ region, both Fig.~\ref{fig:thermospi} and ~\ref{fig:thermospi-lambda0p05}
 show a gradual decrease in the value of $S(\pi)$ as the system undergoes a transition from SLMI to SF phase.

 \begin{figure}[t]
  \centering
  \psfig{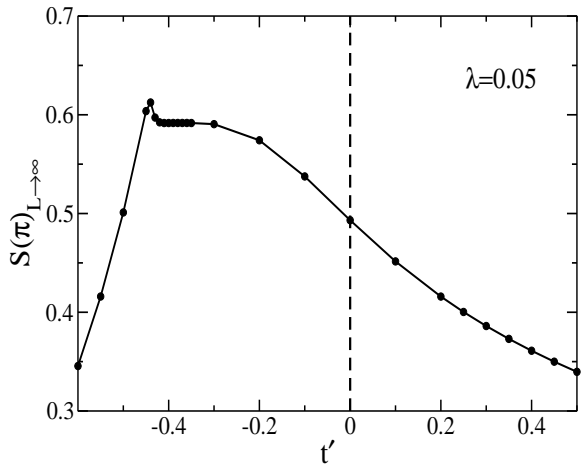}
  \caption{Thermodynamic values of $S(\pi)$ is plotted for the entire range of $t'$ for
  $\lambda=0.05$.}
  \label{fig:thermospi-lambda0p05}
 \end{figure}
 We have also obtained the momentum distribution since it can, in principle, be observed experimentally through time-of-flight images and plotted it in
 Fig.~\ref{fig:momentum}. In BO phase, two peaks appear, which shift away from $k=\pm \pi$. But as the system enters
 the SLMI phase, it shows a broad
 peak at around $k=0$.  (Figs.~\ref{fig:momentum}(a) and (b)).
 The two peak structure in the BO phase is obtained even for $\lambda=0$ and has been investigated earlier~\cite{tapan}.
 For positive $t'$ region, the population of atoms in the $k=0$ state is small, as indicated in
 Fig.~\ref{fig:momentum}(c) but as the system enters the SF region, the $k=0$ starts filling up resulting
 in a large peak as shown in Fig.~\ref{fig:momentum}(d).
 \begin{figure}[!t]
  \centering
  \psfig{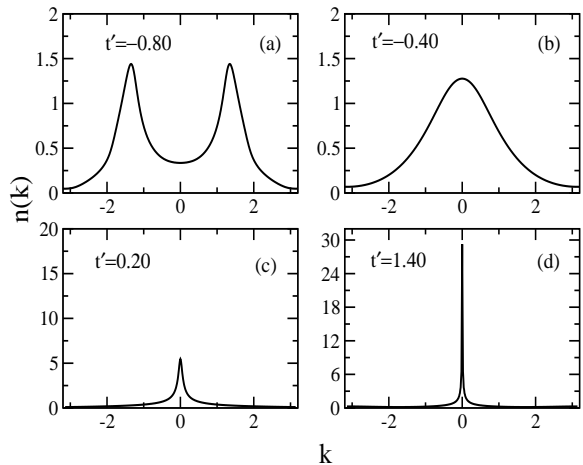}
  \caption{Momentum distribution for different values of $t'$ for $\lambda=0.5$.}
  \label{fig:momentum}
 \end{figure}

\section{Conclusions}
We have obtained the phase diagram of a model of  hard-core bosons
in a 1D lattice with nearest neighbor hopping ($t$ set to the value of $1$)
and next-nearest neighbor hopping ($t'$) in the presence of a
superlattice potential ($\lambda$). 
We find that the phases obtained
depend on the sign of $t'$. For $t'>0$, there are two phases, a
gapped superlattice induced Mott insulator (SLMI) and a gapless superfluid
(SF). The superfluid is stable to switching on the
superlattice potential and a finite value of $\lambda$ is required
to drive the SF into the SLMI. On the other hand for $t'<0$, we
obtain in addition to the SF and SLMI, a gapped bond ordered (BO)
phase. The SF phase for $t'<0$ is unstable to switching on a
superlattice potential and thus exists only for $\lambda=0$ up to a
finite value of $|t'|$. The BO phase exists even when $\lambda =0$
and thus spontaneously breaks lattice translational symmetry and can
be characterized by a bond order parameter. The transition from the
BO to the SLMI phase appears to be first order. 

\section{Acknowledgments}
The computations reported in this work were performed using the
Intel HPC (Hydra) Cluster at IIA, Bangalore, India. We thank Abhishek Dhar for 
many useful discussions. R.V.P.
acknowledges financial support from U.G.C. India. S. M. thanks the Department 
of Science and Technology of the Government of India for support.

\end{document}